\documentclass[preprint,showpacs,preprintnumbers,amsmath,amssymb]{revtex4}

\usepackage{graphicx}
\usepackage{dcolumn}
\usepackage{bm}
\usepackage[latin1]{inputenc}

\begin{document}

\title{Cavity Enhanced Optical Vernier Spectroscopy\\
Broad Band, High Resolution, High Sensitivity}

\author{Christoph Gohle}
\email{ctg@mpq.mpg.de}
\author{Björn Stein}
\altaffiliation[Now at]{ Physikalisch Technische Bundesanstalt,
Braunschweig, Germany}
\author{Albert Schliesser}
\author{Thomas Udem}
\author{Theodor W.\ Hänsch}
 \affiliation{Max-Planck-Institut für Quantenoptik, Hans-Kopfermann-Straße 1,
 D-85748 Garching, Germany}

\date{\today}

\begin{abstract}
A femtosecond frequency comb provides a vast number of equidistantly
spaced narrow band laser modes that can be simultaneously tuned and
frequency calibrated with 15 digits accuracy. Our Vernier
spectrometer utilizes all of theses modes in a massively parallel
manner to rapidly record both absorption and dispersion spectra with
a sensitivity that is provided by a high finesse broad band optical
resonator and a resolution that is only limited by the frequency
comb line width while keeping the required setup simple.
\end{abstract}

\pacs{
42.62.Fi, 
42.62.Eh, 
42.30.Rx, 
42.25.Bs, 
42.30.Ms. 
}
\keywords{Laser, Frequency Metrology, Phase retrival, Cavity ring down, spectroscopy, high resolution, moire patterns}
\maketitle




Femtosecond frequency combs \cite{Udem2002} provide an equidistant
array of narrow line width frequencies that can be simultaneously
tuned. In this paper we demonstrate how to employ such a frequency
comb to record broad band, high sensitivity absorbtion {\em and}
dispersion spectra with potential Hertz resolution in very short
time. The high sensitivity is provided by means of a simple Fabry
Perot resonator that vastly extends the interaction length with the
sample under study and has the potential to achieve sensitivities of
classical ring down techniques. The Hertz resolution is provided by
an appropriately stabilized laser \cite{Bartels2004}. As the
frequency comb can readily be referenced to a primary frequency
standard, the frequency accuracy and reproducibility can be as good
as $10^{-15}$. Such a system could prove useful for quick and
reliable identification of molecular species with complex amplitude
and phase structure. To achieve comb mode resolution while keeping
the setup simple, the resonator acts as a filter that thins out the
frequency comb such that each individual mode can be resolved with
the help of a small diffraction grating. The resonator modes are
scanned across the frequency comb like a Vernier scale in frequency
space. Groups of comb lines that contain the spectroscopic data are
streaked on a two dimensional array. In this way a comparitively
slow detetcor such as a charge coupled device (CCD) can be used to
quickly aquire data. Similar to a previous method
\cite{Schliesser2006} it allows to determine the dispersive details
of the resonator itself which is essential for designing femtosecond
enhancement resonators that have recently been used to generate
frequency combs in the extreme ultraviolet and beyond
\cite{Gohle2005,Jones2005}.

In the last decade a variety of methods were introduced, which
exploit the unique combination of broad bandwidth and high temporal
coherence of a frequency comb for sensing and identifying
(``fingerprinting'') chemical species\cite{Keilmann2004,
Schliesser2005, Schliesser2006, Crosson1999, Gherman2002,
Thorpe2006, Thorpe2007}. Dispersion compensated Fabry Perot
resonators exhibit an equidistant grid of resonances that can be
lined up with the modes of the frequency comb, which has the
consequence that the entire driving power can be simultaneously
resonant. Cavity enhanced absorbtion spectroscopy methods (like
cavity ring down) take advantage of this fact to achieve a higher
signal to noise ratio \cite{Crosson1999, Gherman2002, Thorpe2006,
Thorpe2007}. The resolution of these methods was however limited by
the spectrometer dispersing the light after its passage through the
resonator. Resolving individual comb components on the other hand
can boost the resolution down to the Hertz level that can only be
provided by stabilized lasers \cite{Bartels2004}. Doing so over a
large bandwidth with a grating proves challenging as such a grating
has to be very large and expensive if operated in first order. If
operated in higher order the bandwidth is limited due to overlapping
diffraction orders. An effective and simple method to simultaneously
resolve and record a vast number of individual comb modes has
recently been demonstrated \cite{Diddams2007}. It combines two
spectrally dispersing elements in an arrangement similar to a
prism-echelle spectrometer. One of these operates in high order and
provides comb resolution, while the second operates in first order
to separate the overlapping diffraction orders of the first. This
approach is however difficult to combine with a broadband ring down
technique \cite{Thorpe2006} as both methods require a two
dimensional array for detection.

\begin{figure}[th]
\includegraphics[width=\linewidth]{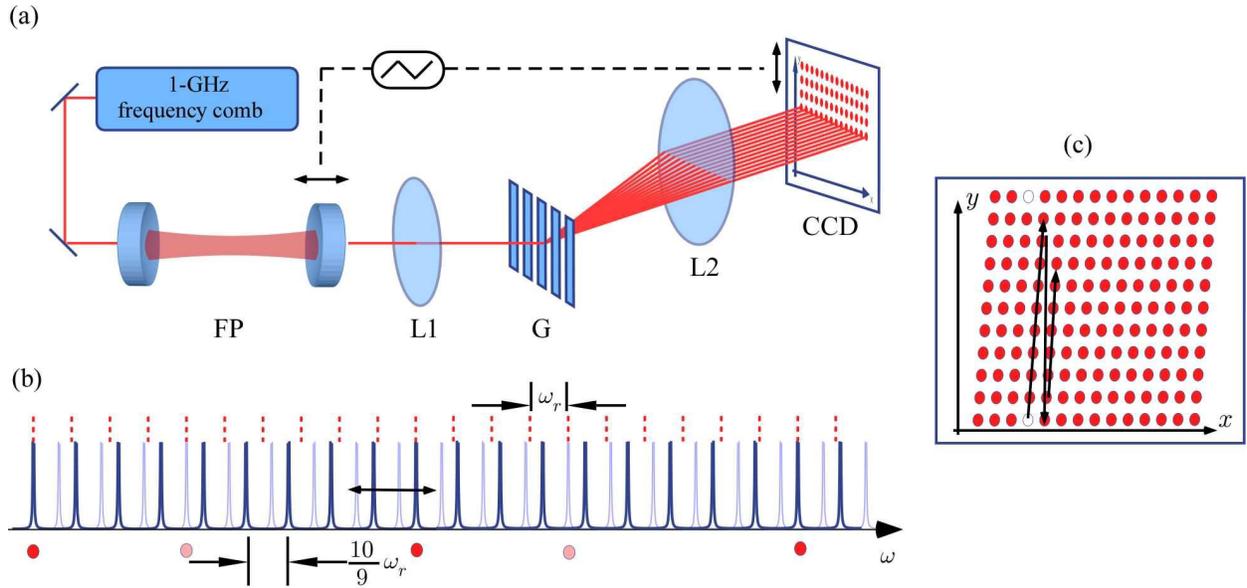}
\caption{\label{setup} a) General setup: A femtosecond laser output
is sent to a Fabry Perot (FP) resonator with a scanning mirror. The
transmitted light is spectrally resolved by a grating spectrograph.
Synchronously with the FP , the CCD is scanned orthogonal to the
spectrograph dispersion to map the transmission spectrum of the FP
as a function of cavity tuning to the y axis on the CCD. L1,L2:
lenses, G: grating. Panel b) frequency comb (dashed) transmission
function of the resonator (solid). The resonator is detuned to the
comb to provide a Vernier for the frequency comb, which is shifted
as the resonator length changes. Dots mark transmitted modes for two
positions of the Vernier. Panel c) Expected pattern on the screen
behind the spectrometer. Arrows mark the order of the frequency comb
modes (index increasing by one from spot to spot along the arrows,
open spots belong to the same comb frequency) }
\end{figure}
In our optical Vernier spectrometer (fig.\ \ref{setup}), this
difficulty is circumvented by combining the comb mode resolving,
high diffraction order part of the echelle spectrometer
\cite{Diddams2007} and the sensitivity enhancing high finesse
resonator into one device. The output of a frequency comb laser has
a spectrum containing only an equidistant grid of frequencies
$\omega_k=\omega_{CE}+k\omega_r$, where $\omega_r$ is the pulse
repetition frequency of the laser, $k$ is a large integer and the
components are offset from being integer multiples of $\omega_r$ by
the carrier envelope offset frequency $\omega_{CE}$. This light is
sent to the high finesse resonator, which contains the sample under
study. It has resonances spaced by its free spectral range FSR$=2\pi
c/nL$, where $c$ is the speed of light in vacuum and $n$ is the
effective refractive index. The geometrical length $L$ of the
resonator is adjusted such that, if the two scales are lined up,
every $m$-th mode from the frequency comb will be resonant with
every $n$-th mode of the cavity and transmitted through it. Provided
the finesse of the resonator is high enough (i.e. larger than $m$),
the other modes will be strongly suppressed. In this way the modes
of the frequency comb and the resonant frequencies of the resonator
resemble a Vernier (fig.\ \ref{setup}b) with the {\em Vernier ratio}
$FSR/f_r=m/n$ being a cancelled fraction. In our case this is chosen
to be $69/68$ ($10/9$ in the figure for clarity). By choosing
$m=n+1$ the interaction length in the resonator is essentially the
same as in a matched resonator, instead of $1/m$ for $n=1$ which
would significantly reduce the sensitivity. Then the frequency comb
that is transmitted through the resonator has a spacing of about
$m\times1$~GHz, so that for sufficiently large $m$ it can easily be
resolved with a very simple and small grating spectrograph. As the
length of the resonator is now tuned, the Vernier primarily shifts,
bringing the next set of $m\times1$~GHz spaced comb modes into
resonance. After scanning the resonator length by half a wavelength
(i.e. one free spectral range), $m$ groups of $m\times1$~GHz combs
have passed by and the situation is as in the beginning, the initial
group is transmitted. The only difference is that the Vernier ratio
has slightly changed due to the change in resonator length.

Now a mirror rotates synchronized with the scanning resonator to
streak the light transmitted through the resonator along the axis
(y-direction) that is perpendicular to the grating dispersion
(x-direction). A 2-dimensional CCD detector is (also synchronized to
the scanning) exposed to the output of that spectrometer (in figure
\ref{setup}a a moving detector is shown for simplicity). This
results in a 2-dimensional array of spots on the CCD that can
uniquely be ordered into a single frequency axis as shown in fig.\
\ref{setup}c, very similar to the sorting procedure used in ref.\
\cite{Diddams2007}. Like this we can associate the comb mode
frequency $\omega_{i,j}=\omega_{k0+i+mj}$ to the spot in row $i$ and
column $j$ (counting from an arbitrary spot $(0,0)$, belonging to
frequency $\omega_{k0}$). Both the precise value of $m$ and $k0$,
and therefore absolute frequency calibration of the data can easily
be determined if there is a sufficiently narrow feature of known
(better than $f_r$) frequency in the spectrum (e.g. an absorbtion
resonance, blue spots in fig.\ \ref{setup}c). This feature will
repeat itself as the resonator is tuned and the period (in spots)
gives the correct value for $m$. Of course because of the spectrum
is sampled at comb frequencies only, the resolution of a single such
pattern is on the order of $f_r$ and aliasing effects may occur. An
essentially arbitrarily high resolution can be achieved by recording
a series of spectra with different center frequencies of the
frequency comb to fill the gaps.

The signal of the spectroscopy is contained in the brightness and
shape of the spots as well as their position. To quantitatively
interpret this pattern, we consider a simple steady state model for
the signal that is observed on the CCD as a function of frequency
(x-axis on the CCD) and resonator length (y-axis on the CCD).
Assuming the individual comb modes can be well isolated from each
other on the CCD image, we can approximate the transmission function
in the vicinity of spot $(i,j)$ as a simple Airy function
originating from a single comb mode $\omega_{i,j}$, i.e.
\begin{equation}\label{eq:transspec}
    \mathcal{T}_{i,j}(x,y)=\frac{T^2
    \delta(\omega(x)-\omega_{i,j})}{1+r(\omega_{i,j})^2-2r(\omega_{i,j})\cos(\phi(L(y),\omega_{i,j}))},
\end{equation}
where $r$ is the round trip amplitude loss factor
$r(\omega)=\sqrt{1-A(\omega)}(1-T)$, $T$ the resonator mirror
transmission, $A(\omega)$ the absorption of the medium inside the
resonator per round trip, $\delta$ the line shape function of the
individual frequency comb mode and $\omega(x)$ and $L(y)$ are
appropriate calibration functions of the spectrometer. The line
shape function $\delta$ may be approximated by a Dirac delta
function for the most practical purposes. The round trip phase shift
$\phi(L,\omega)=\psi(\omega,L)+L\omega/c$ contains a vacuum
contribution $L\omega/c$ and the additional phase shift
$\psi(\omega,L)$ due to dispersion of the mirrors and the medium.
This Airy function (\ref{eq:transspec}) assumes an infinitely large
aperture spectrometer and a point source and hence an infinite
resolution. With the parameters chosen in this demonstration
experiment however, the actual spot shapes are dominated by the
laser beam profile, which can be approximated by a 2-dimensional
Gaussian of fixed width. The convolution of (\ref{eq:transspec})
with this beam profile may be fit to each spot $(i,j)$ in the image
to obtain both $A(\omega_{i,j})$ and $\psi(\omega_{i,j})$ up to
multiples of $2\pi$, provided $y$ can be calibrated absolutely to
the resonator length $L$ ($\omega(x)$ need not be known). We derived
the scaling for $L(y)$ from the fact that $\omega_{i,j}$ and
$\omega_{i+m,j-1}$ represent the same frequency, so that the phase
shift difference between the two spots needs to be $2\pi$ and
therefore $L(y_{i+m,j-1})-L(y_{i,j})=2\pi c/\omega_{i,j}$. Here
$y_{i,j}$ represents the y-position of the maximum of the peak on
the CCD. The dependence of $\psi(L,\omega)$ on the resonator length
was neglected in this expression. This is a good assumption as long
as the extended medium has a refractive index very close to unity
(or has a fixed length). Using this relative length calibration, one
may determine $\psi(\omega)$ up to a linear function.

\begin{figure}
\includegraphics[width=.8\linewidth]{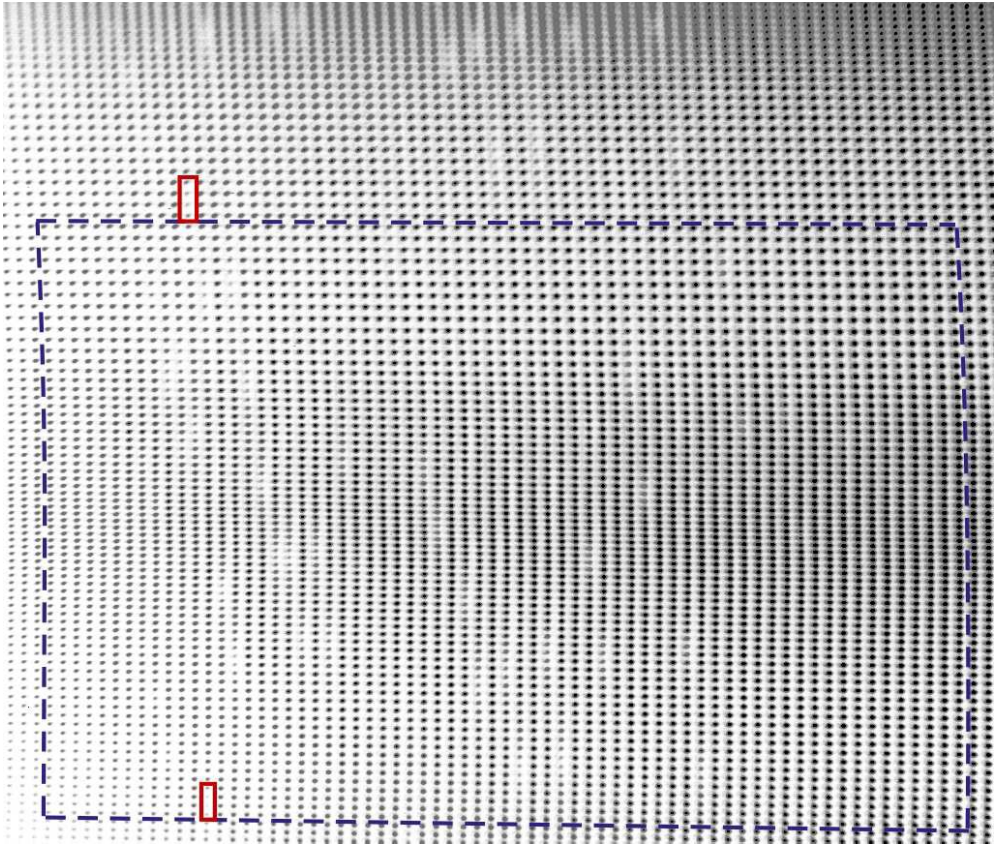}
\caption{\label{data} Raw data image as seen by the CCD, white
corresponds to no light. The horizontal axis is the dispersion axis
of the grating and the vertical axis is the scan axis. Dark spots
caused by absorption lines in the A-bands of O$_2$ are clearly
observed. The period in the pattern in vertical direction identifies
the detuning to 69/68th (solid boxes mark exemplary one pattern that
was identified). The dashed box marks a unique data set. The varying
spot spacing and brightness common to all spot columns originates
from a nonlinearity of the scan and is corrected for in the data
analysis.}
\end{figure}

We use a 1~GHz mode locked Ti:Sapphire laser with a bandwidth of
about 12~THz FWHM centered at 785~nm and an average output power of
about 0.5~W as a light source \cite{Bartels1999}. The spectroscopy
resonator consists of broadband dielectric quarter wave stack
mirrors, centered at 792~nm and arranged in a bowtie ring cavity
configuration. The FSR is about $(69/68)f_r$. The two coupling
mirrors dominate the resonator loss with their nominal transmission
of 0.1\%, corresponding to a resonator finesse of $1000 \pi$, which
we confirmed by ring down measurements with the laser operated in a
continuous mode. One of the coupling mirrors is mounted on a piezo
electric transducer tube for scanning the resonator length. The
buildup time of 1~$\mu$s sets a lower limit of about 3~ms on the
scan time over a full free spectral range in order to satisfy the
steady state approximation that was used in (\ref{eq:transspec}).
We use a cm size 2100 lines per mm holographic grating in a 150~mm
focal length Cherny-Turner arrangement. The aperture of the grating
is fully used to match the resolution of the spectrometer to the
pixel size (5~$\mu$m) of  $1000\times 1200$ pixel CCD camera that
was used to record the images (WinCAM-D 1M4). The resolution of the
spectrometer expected from these paramters is better than 30~GHz at
760~nm.
%
Raw data in the wavelength range from 760 to 770~nm with air inside
the resonator were obtained in the way described above and are shown
in figure\ \ref{data}. This image was exposed for 9~ms, the time for
one scan.

After aquisition, the period of an observed absorption pattern was
used to confirm the detuning ratio of the resonator to be 69/68. To
obtain brightness values for each spot, a 2-d Gaussian was fit to
the the spot and the its integral was used as a brightness value.
This is a good approximation to the model (\ref{eq:transspec}), as
explained above. Because this is the case, the integral was related
to the Airy (\ref{eq:transspec}) area to obtain an intracavity
absorption value. Figure~\ref{spectrum}a shows an absorption
spectrum obtained like this after applying a linear filter to the
data (see below). A comparison with the HITRAN \cite{Rothman2005}
molecular spectroscopic database reveals that the absorption feature
is caused by a $X {}^3\!\Sigma^-_\mathrm{g} \rightarrow b
{}^1\!\Sigma^+_\mathrm{g}$ magnetic dipole intercombination
transition in molecular oxygen \cite{Krupenie1972}. We used this
identification to calibrate the offset frequency $\omega_{CE}$, that
was not measured independently in this demonstration. Note that our
measurement perfectly agrees in frequency scale, amplitude and line
width with the database values (fig.\ \ref{spectrum}).
\begin{figure}
\includegraphics[width=.65\linewidth]{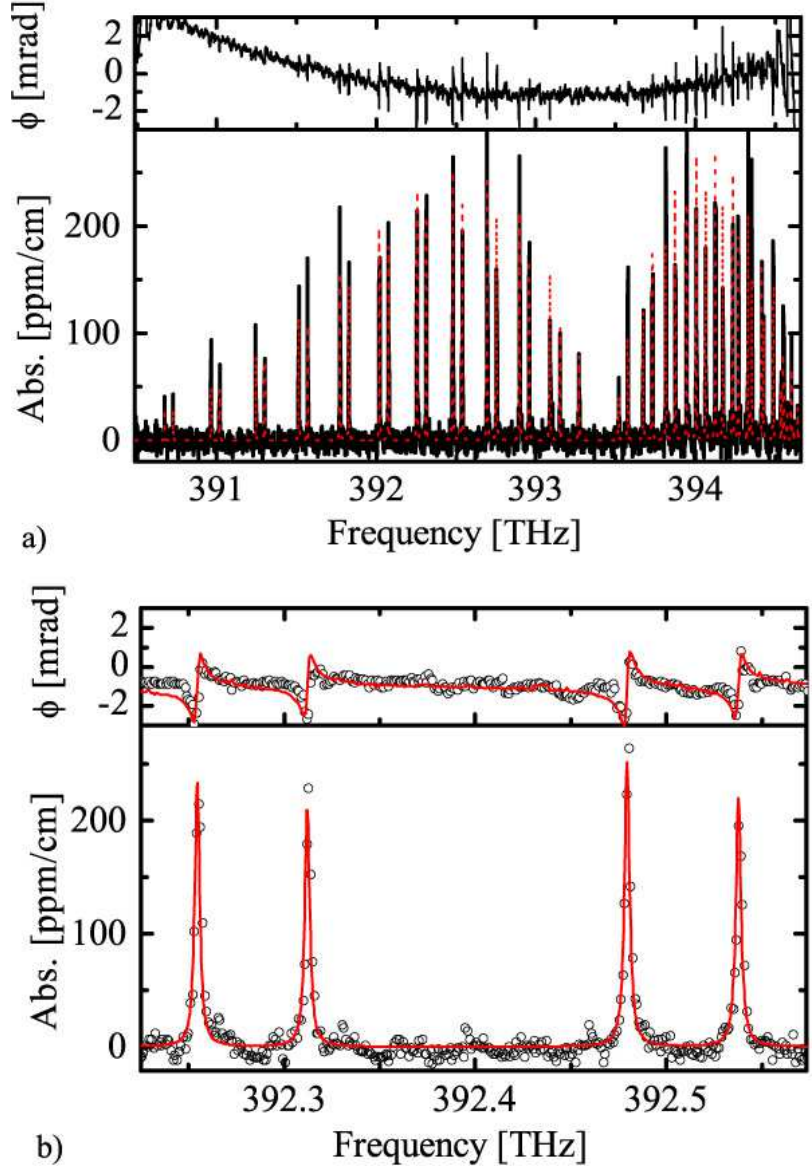}
\caption{\label{spectrum} Panel a) shows the entire spectrum
extracted from the raw data shown in fig.\ \ref{data} (solid line),
which is in good agreement with the HITRAN database values
\cite{Rothman2005} (dashed red line). The phase shows a global
positive second order dispersion of the resonator as compatible with
the mirror design characteristics and spikes at the resonances. The
only free parameter, a global frequency offset, we obtained by
matching one absorption peak to the HITRAN data. Panel b) shows a
zoom of the dataset, highlighting the excellent agreement between
our data (circles) and the HITRAN database (red line).}
\end{figure}

One strength of the method is to provide both amplitude and phase
information about the sample. The phase information that could be
extracted is shown in figure\ \ref{spectrum}. This (round trip)
phase spectrum has a suitable linear function subtracted in order to
make the dispersion more easily be seen. It shows a global second
order dispersion, as expected from the dielectric mirrors that were
used at the blue end of the mirrors' reflectance. On top of that
local dispersive features are seen, which agree, in position and
amplitude, with the expected dispersive features of the O$_2$
resonances. The phase sensitivity with the current implementation is
not specifically high, as the Airy pattern is not resolved by the
spectrometer so that the phase information is degraded.

By evaluating the rms absorbtion noise level at positions where no
signal is detected, we estimate the absorbtion sensitivity of the
method for a single frequency to be $5\times10^{-6}/\mathrm{cm}$ for
a single image taken within 10~ms. With an appropriate camera with
no (small) dead times between acquisitions, this can lead to a
sensitivity of $5\times10^{-7}/\mathrm{cm}\sqrt{\mathrm{Hz}}$
already for this proof-of-principle experiment. Additionally, the
shown spectrum was recorded in the far wing of the laser spectrum
where little power was available and one can expect the signal to
noise to increase with more driving power.

The most severe limitation in this preliminary implementation comes
from acoustic noise and the pizeo transducer's imperfections, which
render the length scan of the resonator mirror nonlinear, leading to
varying vertical positions and brightness of the observed spot
pattern. However, the scan position is the same for any spot with
the same y-value and such distortions will be the same for every
spot column. The higher spot density in the vertical center of the
data fig.\ \ref{data} is an example of such a pattern. An
appropriate linear filter was used to reduce the magnitude of this
artifact and the result is shown in fig.\ \ref{spectrum}. Of course,
this procedure will also reject real structure that by accident has
the same property. Such a case can be tested for by changing the
resonator detuning, which in turn will make the real structure
reappear. Still this and other noise sources reduce the sensitivity
from the shot-noise limited $3\times10^{-9}/\mathrm{cm}$ to the
observed level of $5\times10^{-6}/\mathrm{cm}$ in $10\,\mathrm{ms}$,
and does particularly also degrade the quality of the phase data. In
an improved version, this excess noise can be eliminated by using a
stable optical reference (diode laser, offset locked to the
frequency comb) to which the resonator can be locked. With this and
resolved Airy fringes, it should be possible to considerably reduce
the excess noise on both amplitude and phase outputs.

In conclusion, we have presented a simple method that provides comb
line resolution over a band width of more than 4 THz, which should
be easily extendible to 40~THz by packing the spot pattern more
tightly. Simultaneously the sensitivity is better than
$10^{-6}/\mathrm{cm}\sqrt{\mathrm{Hz}}$ per mode which was achieved
in an interaction length of 29~cm only using a moderately high
finesse optical resonator. The method can be made equivalent to a
ring down method by increasing the scan speed and replacing
(\ref{eq:transspec}) with an appropriate non steady state version
\cite{An1995} in order to get into a real ring down regime. Then, if
the streak speed is chosen appropriately so that the ring down wave
form can be seen, the decay time may be separated from the influence
of technical noise. In addition to ring down methods it also
provides dispersion information with a potential accuracy equivalent
or exceeding the one presented in \cite{Schliesser2006}. Due to the
scanning resonator approach, it is not necessary to accurately
control intra-cavity dispersion, which in other methods results in a
reduced transmission or even spurious signals. The price that has to
be paid for these advantages compared to matched resonators is, that
the average power transmission through the resonator is $1/$finesse
so that the shot noise limited absorption sensitivity only scales as
the square root of the finesse instead of the finesse itself. Yet
our method is fast. A single acquisition takes only about 10~ms
giving 4000 data points (4~THz with a resolution of 1~GHz). With
little optimization, one can easily obtain 20-40 thousand spots on a
mega pixel CCD (the only requirement being that the spots are well
separated). The resolution of the scheme is essentially unlimited
(Hz level, due to potential laser line width). And it can, at any
time, be referenced to a primary frequency standard, giving a
frequency reproducibility and accuracy of $10^{-15}$ (sub-Hertz
level!), which is limited by the primary standard only.
%
The potentially high resolution naturally suggests the application
of the demonstrated technique to some kind of cavity enhanced
nonlinear Doppler free spectroscopy \cite{Cerez1980, Ma1990}.
\begin{acknowledgments}
This research was supported by  the DFG cluster of excellence Munich
Centre for Advanced Photonics \url{http://www.munich-photonics.de}.
\end{acknowledgments}
%
\bibliographystyle{apsrev}

\end{document}